\documentclass[12pt,preprint]{aastex}

\usepackage{amsmath}
\input epsf

\shorttitle{Collinder 419}
\shortauthors{Roberts et al.}

\begin{document}

\received{}
\accepted{}

\title{The Membership and Distance of the Open Cluster Collinder 419}

\author{Lewis C. Roberts, Jr.\altaffilmark{1}, 
Douglas R. Gies\altaffilmark{2}, 
J. Robert Parks\altaffilmark{2}, 
Erika D. Grundstrom\altaffilmark{3,8}, 
M. Virginia McSwain\altaffilmark{4,8},
David H. Berger\altaffilmark{5},  
Brian D. Mason\altaffilmark{6},
Theo A. ten Brummelaar\altaffilmark{7}, 
and \\ Nils H. Turner\altaffilmark{7}}

\altaffiltext{1}{Jet Propulsion Laboratory, California Institute of Technology, 
4800 Oak Grove Drive, Pasadena CA 91109; lewis.c.roberts@jpl.nasa.gov}

\altaffiltext{2}{Center for High Angular Resolution Astronomy, 
Department of Physics and Astronomy, Georgia State University, P.O. Box 4106, Atlanta, GA 30302-4106; 
 gies@chara.gsu.edu, parksj@chara.gsu.edu}

\altaffiltext{3}{Physics and Astronomy Department, Vanderbilt University, 6301 Stevenson Center, 
Nashville, TN 37235; erika.grundstrom@vanderbilt.edu} 

\altaffiltext{4}{Department of Physics, Lehigh University, 
16 Memorial Drive East, Bethlehem, PA 18015; mcswain@lehigh.edu}

\altaffiltext{5}{System Planning Corporation, 3601 Wilson Blvd, Arlington, VA  22201; dberger@sysplan.com}

\altaffiltext{6}{US Naval Observatory, 3450 Massachusetts Avenue, NW, Washington, DC 20392-5420; 
bdm@usno.navy.mil}

\altaffiltext{7}{Center for High Angular Resolution Astronomy, Georgia State University, Mt.\ Wilson, CA 91023; 
theo@chara-array.org, nils@chara-array.org}

\altaffiltext{8}{Visiting Astronomer, Kitt Peak National Observatory,
National Optical Astronomy Observatory, operated by the Association
of Universities for Research in Astronomy, Inc., under contract with
the National Science Foundation.}

\slugcomment{Accepted to the Astronomical Journal}

\paperid{}


\begin{abstract}

The young open cluster Collinder~419 surrounds the massive 
O star, HD~193322, that is itself a remarkable multiple 
star system containing at least four components. 
Here we present a discussion of the cluster distance based 
upon new spectral classifications of the brighter members, 
$UBV$ photometry, and an analysis of astrometric and photometric data 
from the UCAC3 and 2MASS catalogs.  We determine an average cluster reddening 
of $E(B-V)=0.37 \pm 0.05$ mag and a cluster distance of $741 \pm 36$~pc.  
The cluster probably contains some very young 
stars that may include a reddened M3~III star, IRAS~20161+4035.  
\end{abstract}

\keywords{stars: individual (HD~193322, IRAS~20161+4035) ---
stars: early-type ---
open clusters and associations: individual (Collinder 419)}

\setcounter{footnote}{8}


\section{Introduction}  

Collinder 419 is an open cluster in the constellation Cygnus.  
It was first designated as Barnard~794 (Barnard 1927), 
but the accepted name comes from the work of 
Collinder (1931) who classified the cluster as a $\mu$~Normae cluster.  
This is an intermediate type of grouping between multiple star systems 
and open clusters.  The prototype cluster NGC~6169 contains the 
bright star $\mu$~Normae = HD~149038 and a host of 
surrounding fainter stars.  In the case of Collinder 419, the bright 
central star is HD~193322 = HR~7767 = HIP~100069 = WDS~20181+4044. 
HD~193322 is a multiple star system with a close visual companion
STF 2666 AB (separation of $2\farcs9$ and probably orbitally bound) 
and two more distant companions, STF 2666 AC ($34\farcs0$) and 
TAR 5 AD ($49\farcs6$), which, if bound have very long orbital periods.   
McAlister et al.\ (1987) discovered that the A component is a speckle binary, 
CHR 96 Aa,Ab.  An approximately 31 year orbital period for this pair was computed by 
Hartkopf et al.\ (1993).  In addition, one of the components of the speckle binary 
is a spectroscopic binary with a 311.03$\pm$0.25 
day period (Fullerton 1990; McKibben et al.\ 1998).   
Fullerton (1990) also obtained a spectrum of the B component that suggests 
that it may also be a spectroscopic binary.  Thus, the central AB pair 
may contain as many as five individual stars.   

The primary of the system, Aa1, has a spectral type of O9~V:((n)) (Walborn 
1972), with the suffix indicating slightly broadened lines.  McKibben et 
al.\ (1998) tentatively identify the close spectroscopic companion 
Aa2 and the B component all as B-type stars.  
The magnitude difference of the speckle pair Aa,Ab is $\triangle V = 1.2\pm0.5$ 
(Mason et al.\ 2009) which suggests that Ab is also an early B-type star. 
Both the Aa2 and Ab companions offer us the means to determine mass for the O star 
component Aa1 through measurement of the orbital motion.  
By themselves, neither an orbit of the Aa,Ab pair nor the spectroscopic 
Aa1,Aa2 pair will yield a mass, but, in conjunction 
with a distance determination, a mass can be determined.  
The \textit{Hipparcos} Catalogue (Perryman et al.\ 1997) has become the 
standard source for parallaxes, but it may be unreliable for some double stars 
(Shatskii \& Tokovinin 1998) and also for distant O stars (Schr\"{o}der et al.\ 2004). 
Until a new space-based parallax engine is available sometime in the next
decade we are left with classical techniques as the best way to determine 
O-star distances.

We are currently pursuing efforts to determine the orbits of the central 
triple through speckle and optical long baseline interferometry (Turner et al.\ 2008).  
Here we present a new assessment of the distance to the cluster 
that we will use in the forthcoming orbit and mass determination.  
In \S2 we present spectral classifications for the brighter 
members, and we show fits of their spectral energy distributions 
that provide estimates of reddening and angular size. 
We then describe in \S3 an astrometric and photometric study 
of cluster membership based upon data in the UCAC3 catalog 
(Zacharias et al.\ 2010) and the 2MASS catalog (Cutri et al.\ 2003).  
We also present new $UBV$ photometry of the fainter cluster members in \S4.
Our results are summarized in \S5. 


\section{Spectroscopy and Reddening of the Brighter Stars}  

\begin{figure}
\begin{center}
{\includegraphics[angle=90,height=12cm]{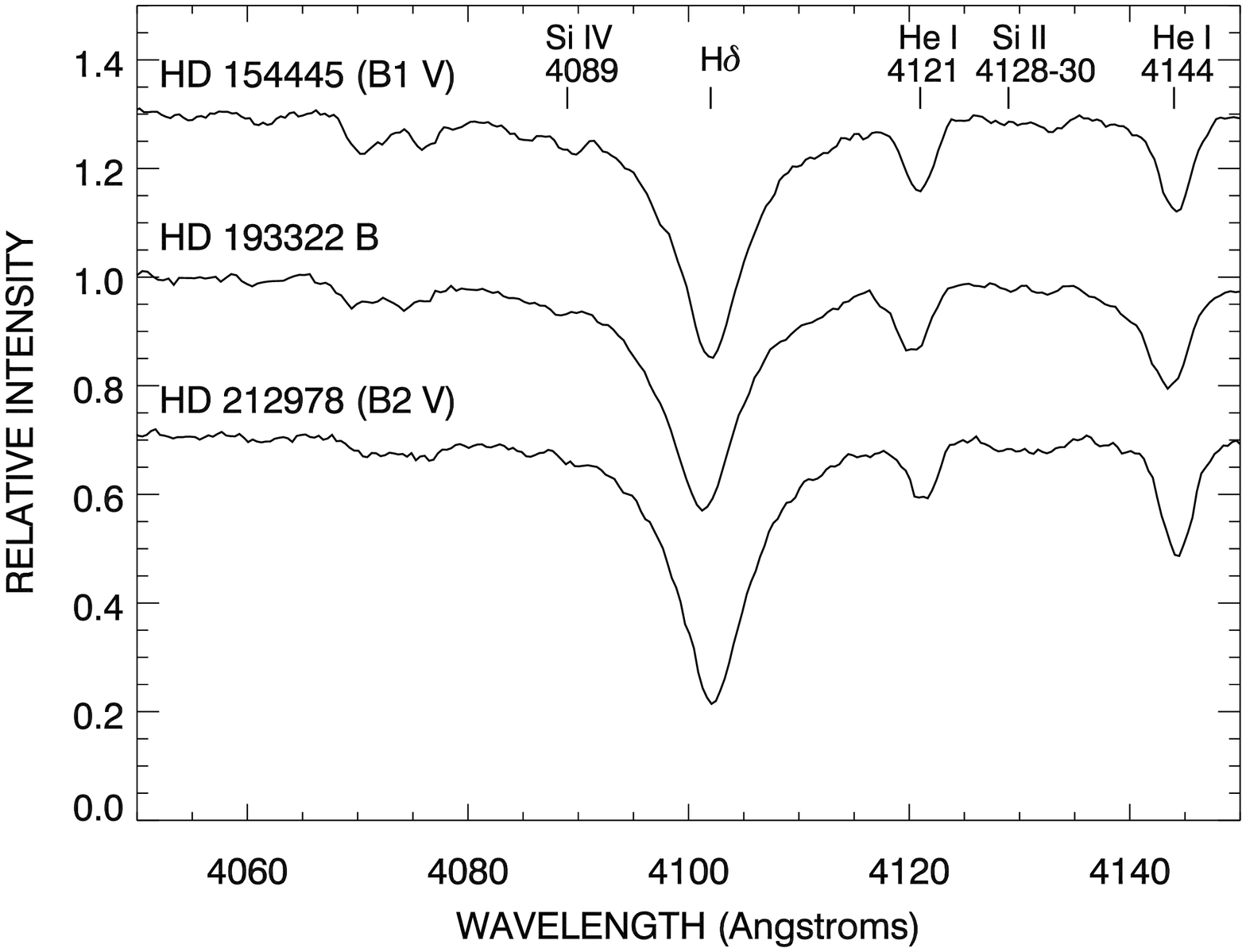}}
\end{center}
\caption[]{Our estimate for the spectral type of HD 193322 B is B1.5 V based on comparisons to stars in \citet{val04}.  \citet{walborn1990} advocate using the Si lines to classify early B stars, and those lines were given the strongest weight.  The spectra were normalized and then offset to improve clarity. 
\label{fig_hd193322B}} 
\end{figure}

\begin{figure}
\begin{center}
{\includegraphics[angle=90,height=12cm]{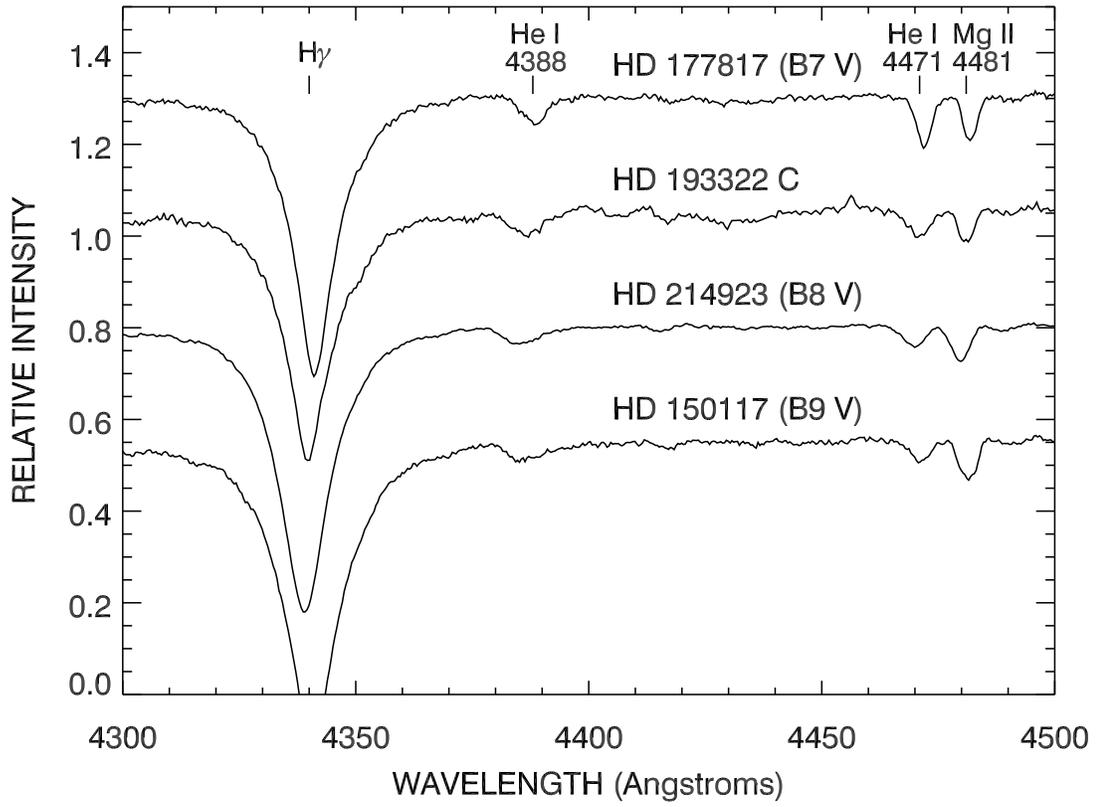}}
\end{center}
\caption[]{ We estimate the spectral type of HD 193322 C as B8 V or possibly B9 V. The spectra were normalized and then offset to improve clarity. 
\label{fig_hd193322C}} 
\end{figure}

\begin{figure}
\begin{center}
{\includegraphics[angle=90,height=12cm]{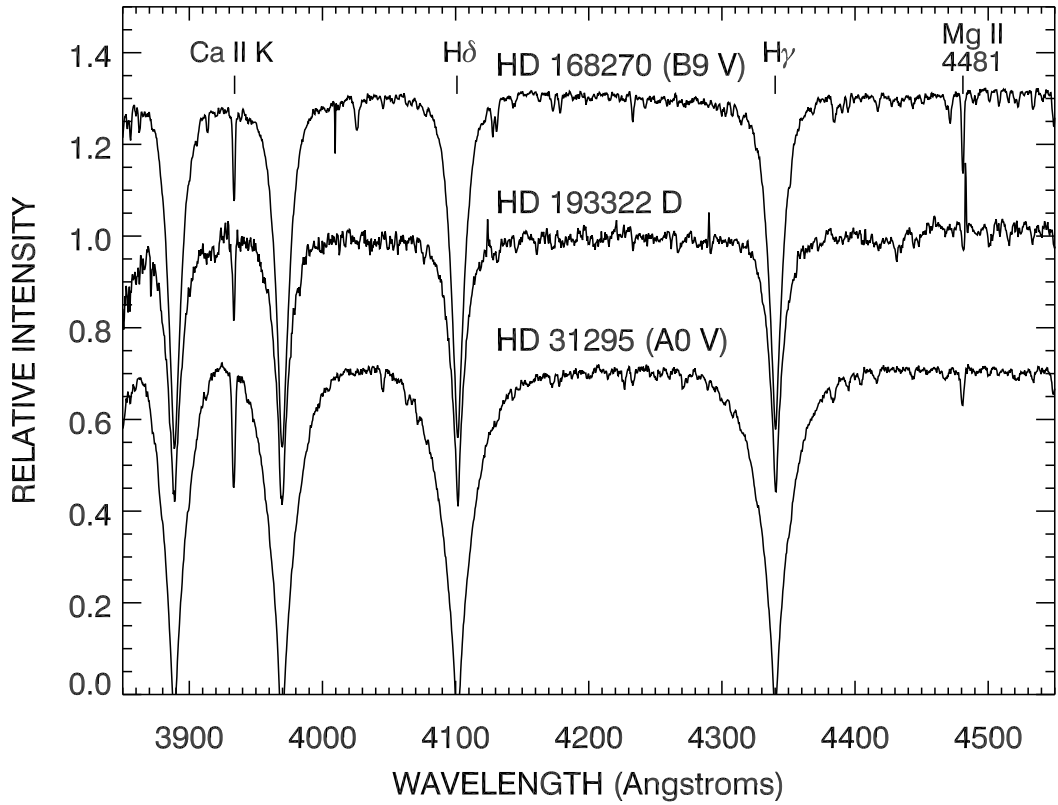}}
\end{center}
\caption[]{We estimate the spectral type of HD 193322 D as B9 V due to the strength of the \ion{Ca}{2} K lines. The H Balmer, He, and metal lines are weak.   The spectra were normalized and then offset to improve clarity. 
\label{fig_hd193322D}} 
\end{figure}

We obtained spectra for five of the brighter stars in the cluster 
in order to determine spectral classifications and to estimate the 
intrinsic colors of the stars.  One of us (MVM) obtained spectra 
of the four blue stars of the central multiple system 
using the Kitt Peak National Observatory (KPNO) 2.1~m telescope 
and Cassegrain focus, GoldCam CCD Spectrograph.   These spectra 
were made during 2005 October and November with the No.\ 47 grating 
(831 grooves mm$^{-1}$) in second order to obtain a resolving power 
$\lambda / \delta \lambda \sim 3000$.  The observed wavelength 
range was 4050 -- 4950 \AA , a region that includes numerous H Balmer, 
\ion{He}{1}, and \ion{He}{2} lines in O- and B-type spectra. 
The data were reduced using standard routines in IRAF\footnotemark
\footnotetext{IRAF is distributed by the National Optical Astronomical
Observatory, which is operated by the Association of Universities for
Research in Astronomy, Inc. (AURA), under cooperative agreement with the
National Science Foundation.}. 
The flux rectified spectra were compared directly to spectra from the 
atlas of Valdes et al.\ (2004) in order to estimate the spectral 
classification based upon the strengths of the H Balmer, 
\ion{Ca}{2}, \ion{He}{1}, and \ion{Mg}{2} lines.  Our results are given in 
Table~1 for the B, C, and D components of HD~193322. See Figures \ref{fig_hd193322B} -- \ref{fig_hd193322D} for the comparision between the target stars and the template spectra.   Our spectra for the 
combined flux of Aa1, Aa2, and Ab agree with the O9~V:((n)) classification
from Walborn (1972), which is listed in Table~1.  These classifications
all agree within a subtype with earlier results (Burnichon 1975; 
Hoffleit 1982; Abt \& Cardona 1983).   

\placetable{tab1}      

\begin{figure}
\begin{center}
{\includegraphics[angle=90,height=12cm]{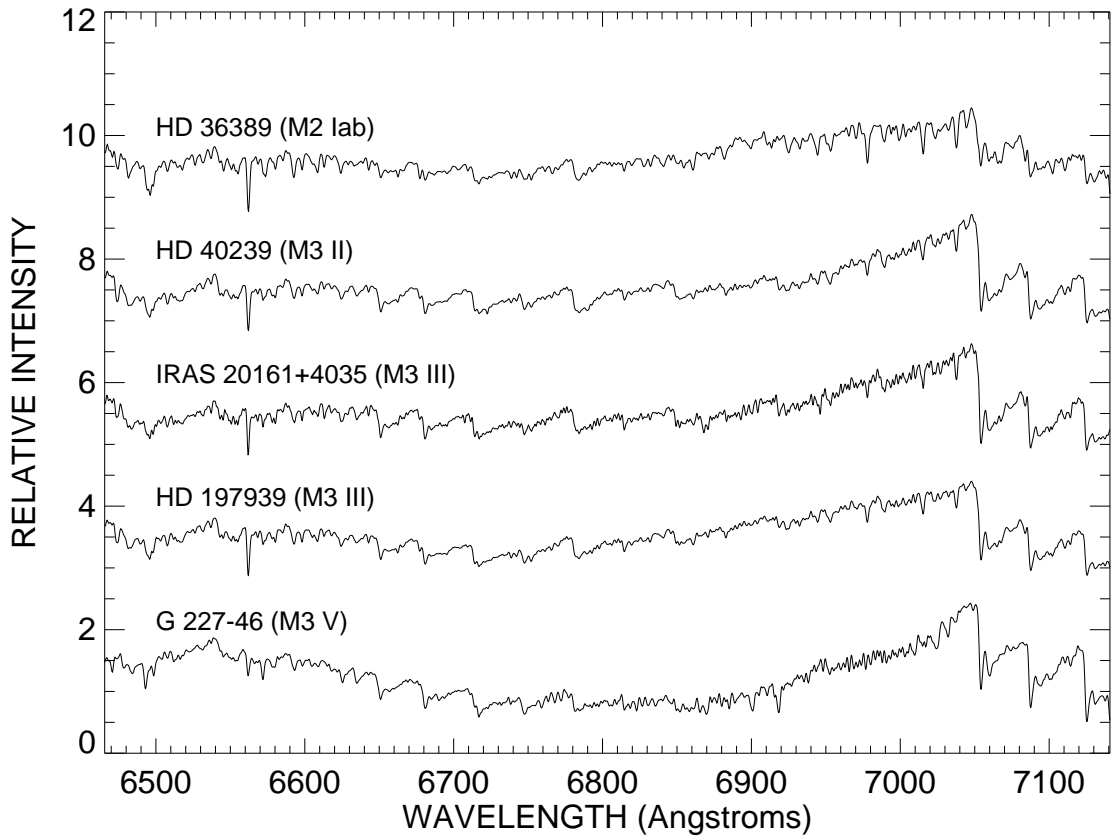}}
\end{center}
\caption[fig1]{A comparison of the red spectrum of IRAS 20161+4035 with 
similar spectra of cool stars from the atlas of Valdes et al.\ (2004). 
Each spectrum has been offset in relative intensity for clarity of 
presentation.
\label{red_spectra}} 
\end{figure}

The very red star IRAS~20161+4035 is the brightest 
object in the $K_s$-band in this vicinity, 
and we were curious to determine if indeed it is related to the 
cluster.  One of us (EDG) obtained three spectra of the object with 
the KPNO 0.9~m Coude Feed Telescope in 2004 October. 
These spectra cover the range from 6460 to 7140 \AA ~with 
a resolving power $\lambda / \delta \lambda \sim 9700$.
The average spectrum is compared 
to several other cool star spectra from Valdes et al.\ (2004) 
(also made with the KPNO Coude Feed telescope) in Figure \ref{red_spectra}. 
We made a digital comparison with all the M2 -- M5 spectra in 
the Valdes et al.\ spectral atlas, and the best fit (based 
mainly on the appearance of the strong TiO bands in this 
spectral region) was the spectrum of the M3~II star HD~40239
(although there were a number of other M3~III spectra which 
made an almost equally good fit).  Figure \ref{red_spectra} shows the good 
overall match made with the M2~II and M3~III giants compared to 
the dwarf and supergiant spectra.  We also measured the H$\alpha$ 
equivalent width in these and similar stars to help establish 
the luminosity class (Eaton 1995).   We found that among stars 
of similar spectral type that the mean equivalent width of 
H$\alpha$ ranged from $0.76\pm0.04$ \AA ~for 2 main sequence stars,
to $1.59\pm0.08$ \AA ~for 4 giant stars, and up to 
$1.89\pm0.04$ \AA ~for 3 supergiant stars.  The measured 
H$\alpha$ equivalent width of IRAS~20161+4035, 1.38 \AA , 
clearly places the star among those in the giant luminosity class,
and we adopted this assignment in Table~1.
Note that the spectrum of IRAS~20161+4035 shows the strongest 
\ion{Li}{1} $\lambda 6707$ feature among the spectra illustrated
in Figure \ref{red_spectra}. 

\placefigure{fig1}     

We can use these spectral classifications to estimate 
the intrinsic colors of the targets, compare these with 
observed multi-color observations, and find the 
reddening $E(B-V)$ in each case.  However, the available 
near-IR 2MASS photometry for HD~193322 corresponds to the
total flux for the entire Aa1, Aa2, Ab, and B complex. 
Thus, we need to make some assumptions about the relative
flux contributions of each component in each filter band
to derive magnitudes and fluxes for the individual stars. 
The magnitude differences of the AB pair 
were determined in the Johnson $BVRI$ bands through 
adaptive optics imaging by ten Brummelaar et al.\ (2000).
Unfortunately, much less is known about the A sub-components. 
Mason et al.\ (2009) found a magnitude difference
of $\triangle V = 1.2\pm0.5$ mag for the Aa,Ab speckle pair, 
and McKibben et al.\ (1998) used a statistical method to
estimate the mass ratio and magnitude difference 
$\triangle V = 1.3$ mag for the Aa1,Aa2 pair in the
spectroscopic binary.   These magnitude differences 
(plus $\triangle V$ for the AB pair from ten Brummelaar 
et al.\ 2000) lead to $V$-band flux ratios relative
to star Aa1 that are given in Table~2.   To derive the 
flux ratios in other bands, we estimated their colors 
according to their spectral classification and the 
color calibration from Wegner (1994). 
We assumed that the Aa2 and Ab stars are
main sequence objects and then estimated their spectral 
types (B1.5~V and B1~V, respectively) based on their magnitude
differences relative to Aa1 (using the absolute magnitude versus 
spectral type relation for main sequence stars from Lesh 1979).  
Table~2 lists the adopted flux 
ratio estimates for the sub-components (which agree with 
those from ten Brummelaar et al.\ for the Johnson $R,I$ bands,
but not for their $B$-band result for which component 
B is unusually faint in $B$ relative to the other bands). 
We used these flux ratios to determine the magnitudes of 
Aa1 and B, the stars with reliable spectral classifications,
in the following analysis.  Note that the current flux ratio 
uncertainties probably introduce an error of $\pm0.2$ mag 
into the error budget for the magnitudes of these two stars. 

\placetable{tab2}      

\begin{figure}
\begin{center}
{\includegraphics[angle=90,height=12cm]{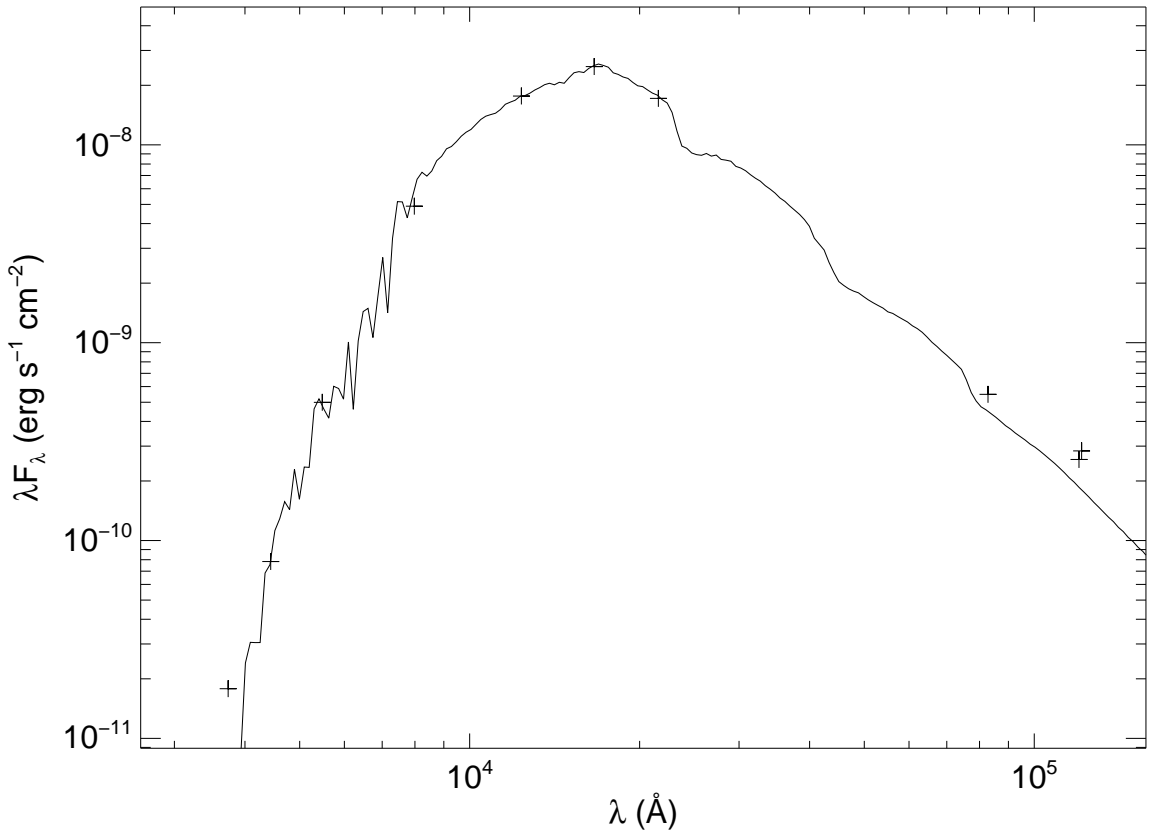}}
\end{center}
\caption[fig2]{The spectral energy distribution of IRAS20161+4035 
({\it plus signs}) compared with a reddened model SED ({\it solid line}).
\label{fig_sed}} 
\end{figure}

We estimated the reddening and angular sizes of the five 
bright stars listed in Table~1 by comparing their observed 
fluxes with reddened model spectral energy distributions. 
The optical magnitudes were taken from the work of 
ten Brummelaar et al.\ (2000; for components Aa1 and B), 
Burnichon \& Garnier (1976; for components C and D)
and Droege et al.\ (2006; plus $UBV$ magnitudes given in \S4 for IRAS 20161+4035), 
and these were combined with 2MASS magnitudes for the near-IR 
(Cutri et al.\ 2003).   The magnitudes were transformed
to fluxes using calibrations from Colina et al.\ (1996; Johnson $U,B,V$), 
Bessell et al.\ (1998; Cousins $I_c$), and Cohen et al.\ (2003; 2MASS $J, H, K_s$).
We used Kurucz model atmospheres\footnote{http://kurucz.harvard.edu/} 
(assuming $\log g = 4.0$, solar abundances, and a microturbulent
velocity of 2 km~s$^{-1}$) for temperatures from 
the spectral calibration of B\"ohm-Vitense (1981; see Table~1) to create 
low resolution versions of the unreddened spectral energy distributions (SEDs). 
However, for the case of IRAS 20161+4035 (M3~III), we adopted a flux 
distribution from the MARCS code\footnote{http://marcs.astro.uu.se/}
(Gustafsson et al.\ 2008) for a model with $T_{\rm eff} = 3500$~K, $\log g =1.0$, 
solar abundances, and a microturbulent velocity of 5 km~s$^{-1}$. 
We then fit the observed fluxes 
with model SEDs reddened according to the formulation from 
Fitzpatrick (1999).  Because the stars have low reddening and limited  
short-wavelength coverage, it is difficult to obtain an independent 
estimate of the ratio of total-to-selective extinction $R$, so we 
adopted the default value $R=3.1$ throughout.  The derived 
reddening $E(B-V)$ and limb-darkened angular diameter $\theta_{LD}$ 
are given in Table~1.  The blue stars yield a consistent reddening
of $E(B-V)=0.29\pm0.01$ mag, which is significantly less than that found 
for IRAS 20161+4035, $E(B-V)=0.74\pm0.04$ mag.  The SED for this 
red star is shown in Figure \ref{fig_sed}, and we see that there appears to 
be an IR flux-excess of $48\pm10\%$ near $12\mu$m, based upon 
data from the MSX (Egan \& Price 1996) and 
IRAS\footnote{Infrared Astronomical Satellite Catalogs, 1988; 
The Point Source Catalog, version 2.0, NASA RP-1190} missions.  
We argue below (\S5) that the higher reddening and IR-excess may be 
related to the youth of this star. 

\placefigure{fig2}     
  

\section{Astrometric and Photometric Data from UCAC3 and 2MASS}  

\begin{figure}
\begin{center}
{\includegraphics[angle=90,height=12cm]{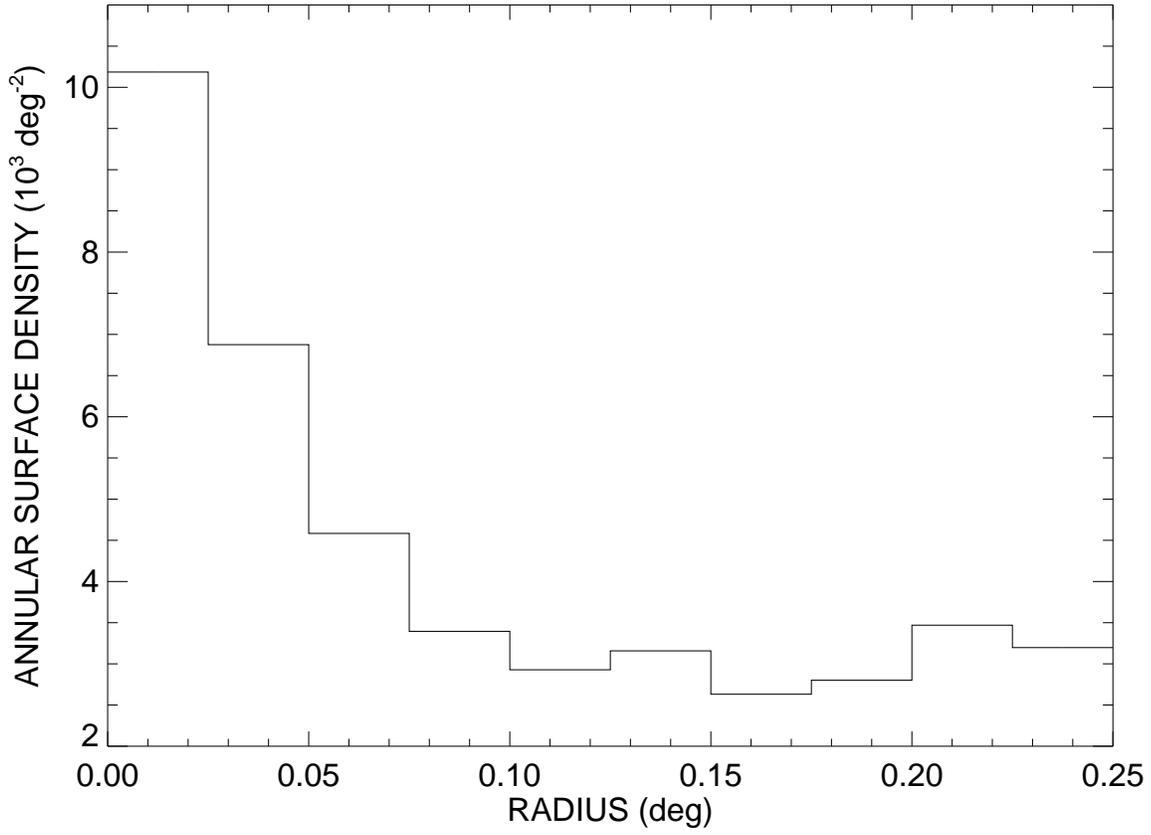}}
\end{center}
\caption[fig3]{The distribution of the stellar number density per unit area
in rings of increasing radius centered on HD~193322, the nominal center of Cr~419.
\label{fig_stellar_density}} 
\end{figure}

\begin{figure}
\begin{center}
{\includegraphics[angle=90,height=12cm]{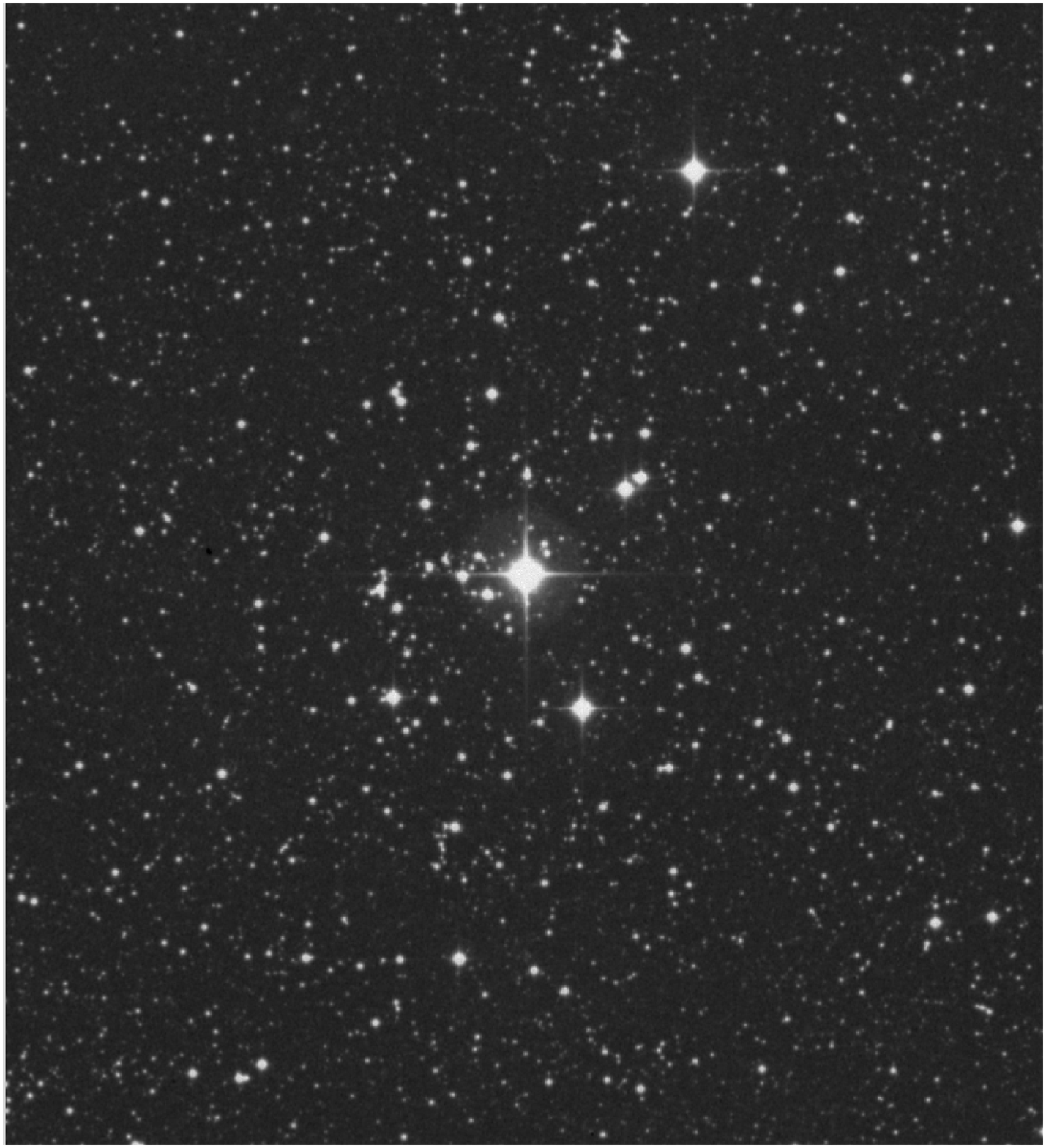}}
\end{center}
\caption[fig4a]{The POSS2-IR image centered on HD 193322 with a field of view of a 0.25$^\circ\times0.25^\circ$ showing the low stellar density of the cluster.  North is up and east is to the left. 
\label{dss_image}} 
\end{figure}

\begin{figure}
\begin{center}
{\includegraphics[angle=90,height=12cm]{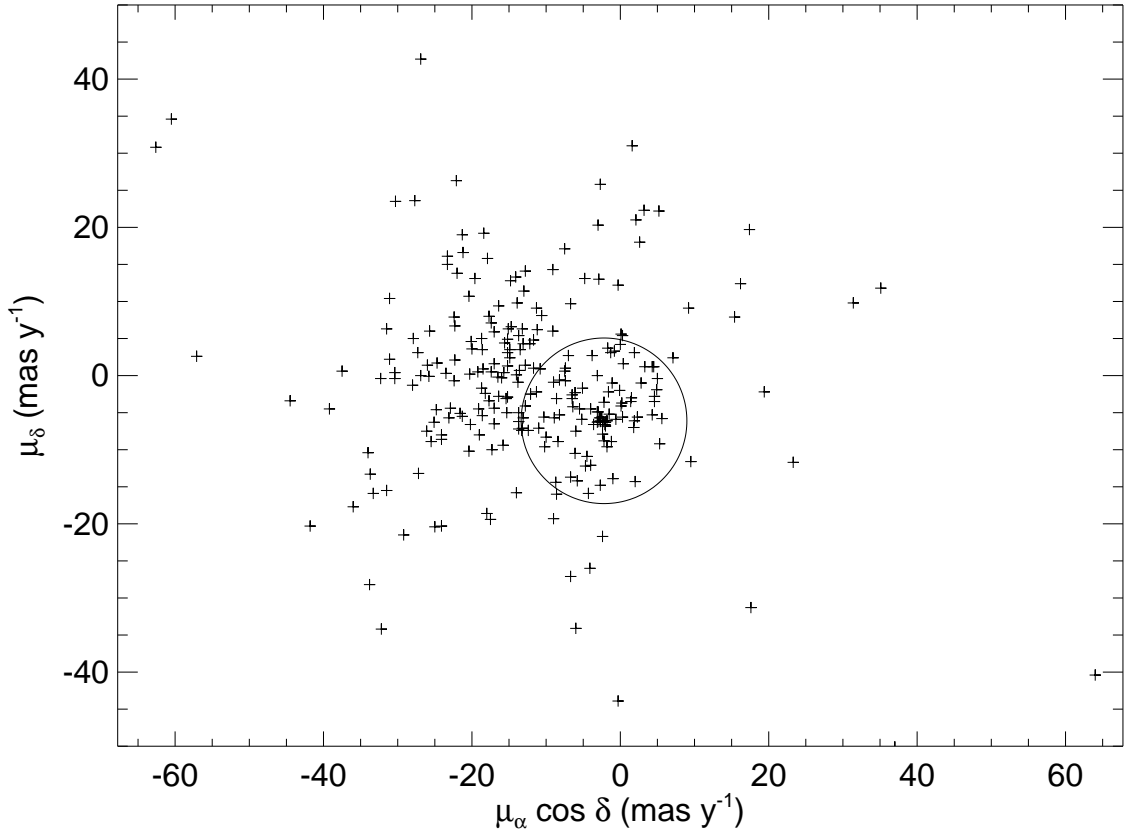}}
\end{center}
\caption[fig4]{The distribution of proper motions in right ascension ($\mu_\alpha \cos \delta$)
and declination ($\mu_\delta$) for stars within $0\fdg16$ of the cluster center.
The position of the circle marks the cluster average proper motion and its
circumference represents the boundary for possible cluster membership.
\label{fig_proper_motion}} 
\end{figure}

With the recent release of the third U.\ S.\ Naval Observatory CCD 
Astrograph Catalog\footnote{http://www.usno.navy.mil/USNO/astrometry/optical-IR-prod/ucac} 
(UCAC3; Zacharias et al.\ 2010), we now have access
to a large collection of astrometric and photometric data 
that can be used to search for additional cluster members.  
We began by selecting all the stars in UCAC3 with complete sets of 
proper motion and photometric data that are found within 
a $0\fdg25$ radius of HD~193322, the nominal center of Cr~419. 
We show in Figure \ref{fig_stellar_density} the average surface density of star counts 
within annuli centered on the position of HD~193322.  
The larger surface density associated with the cluster 
extends outwards to at least $0\fdg10$, and we set the outer 
boundary on cluster membership at $0\fdg16$ to avoid including
too many field stars at large distance.  Figure \ref{dss_image} shows a $0\fdg25$ field centered on HD 193322 and shows the low stellar density of the cluster. The image is from the near-IR plates of Second Palomar Observatory Sky Survey (POSS2-IR) \citep{reid1991}.  Figure \ref{fig_proper_motion} shows the 
distribution of proper motions of stars found within this radius. 
The circle in the diagram is centered on $\mu_\alpha \cos \delta 
= -2.2$ mas y$^{-1}$ and $\mu_\delta = -9.1$ mas y$^{-1}$, 
the average values for the bright cluster stars (HD~193322A, C, D, 
and HD~228810).  There appears to be a bimodal distribution in 
proper motion in this direction, so we selected only those stars 
within the circle, i.e., those with a combined proper motion 
difference from the cluster mean 
of less than 11 mas y$^{-1}$ (the RMS of the
UCAC3 proper motion errors and the expected 
range within the cluster itself).  The constraints of 
position and proper motion reduced the potential sample
to a group of 96 stars. 

The histogram in Figure~\ref{fig_stellar_density} also demonstrates that there
will be a significant number of field stars in this sample.
To estimate this number, we applied the same proper motion
constraint to the UCAC3 sample within a separation range
of $0\fdg16$ to $0\fdg24$, and this indicates that the
surface density of field stars in this direction with
similar proper motions to the bright cluster stars is
$1000\pm100$ stars per $\deg^2$.  Thus, we might expect
to find some $83\pm8$ field stars in the $0\fdg16$ radius
circle centered on HD~193322.  This exercise suggests
that the actual number of cluster stars is small (perhaps
$96$ total - $83$ field $=13$ cluster stars) and that the
sample will be dominated by other field stars.

\placefigure{fig3}     

\placefigure{fig4a}     

\placefigure{fig4}     

We next considered the photometric data in UCAC3, 
which consists of 2MASS photometry and the UCAC3 red-band photometry
that we will refer to as $R_u$ (to distinguish it from the 
Johnson and Cousins counterparts).  
This filter has a bandpass from 579 to 642~nm that falls between
the Johnson $V$ and Cousins $R_c$ bands.  The UCAC3 catalog
reports magnitudes from centroid fit models and aperture photometry, 
and we used the latter in our analysis.  
The UCAC3 listing for HD~193322 corresponds to the total flux from 
components Aa1, Aa2, Ab, and B, and we used the flux ratios given 
in Table~2 to determine the corresponding magnitudes for stars 
Aa1 and B only.   Our goal is to compare the magnitudes of the candidate
member stars with model predictions for given assumptions about the 
reddening $E(B-V)$ and distance modulus DM, in order to find values
of these parameters that lead to the best fit of the photometry.  

We adopted model isochrones for the cluster from the evolutionary 
models of Bertelli et al.\ (1994) and 
Marigo et al.\ (2008)\footnote{http://stev.oapd.inaf.it/cmd}.   
We selected an isochrone for solar metallicity and a cluster age of 7~Myr, 
the value proposed by Kharchenko et al.\ (2005) and one that is consistent 
for a time of flight from Cr~419 of the runaway star 68~Cyg 
(Schilbach \& R\"{o}ser 2008).  The isochrones were produced for a 
number of photometric bands including Johnson $V$, Cousins $R_c$, 
and 2MASS $J, H, K_s$.  

We need to transform the model optical magnitudes into the $R_u$ 
system in order to compare them with the UCAC3 values. 
We used a sample of 402 photometric standard stars from 
Landolt (1992) that appear in UCAC3 to arrive at a transformation 
equation from Johnson $V$ and Cousins $R_c$ to UCAC3 $R_u$,
\begin{equation}
R_u = R_c + 0.292 + 0.372 (V-R_c) + 0.054 (R_c - 13) 
\end{equation}
which successfully fits the observed UCAC3 $R_u$ magnitudes 
over the range $R_u= 8 - 17$ with a standard deviation of 0.116 mag. 
The color term in the equation shows how the $R_u$ magnitude 
changes with stellar color because of the difference in the 
effective wavelengths of the $R_u$ and $R_c$ bands. 
The final term on the right hand side reflects a small systematic
difference between the Landolt and UCAC3 systems that is 
approximately equal to the scatter in the relation at the 
bright and faint ends of the sample.

\begin{figure}
\begin{center}
{\includegraphics[angle=90,height=12cm]{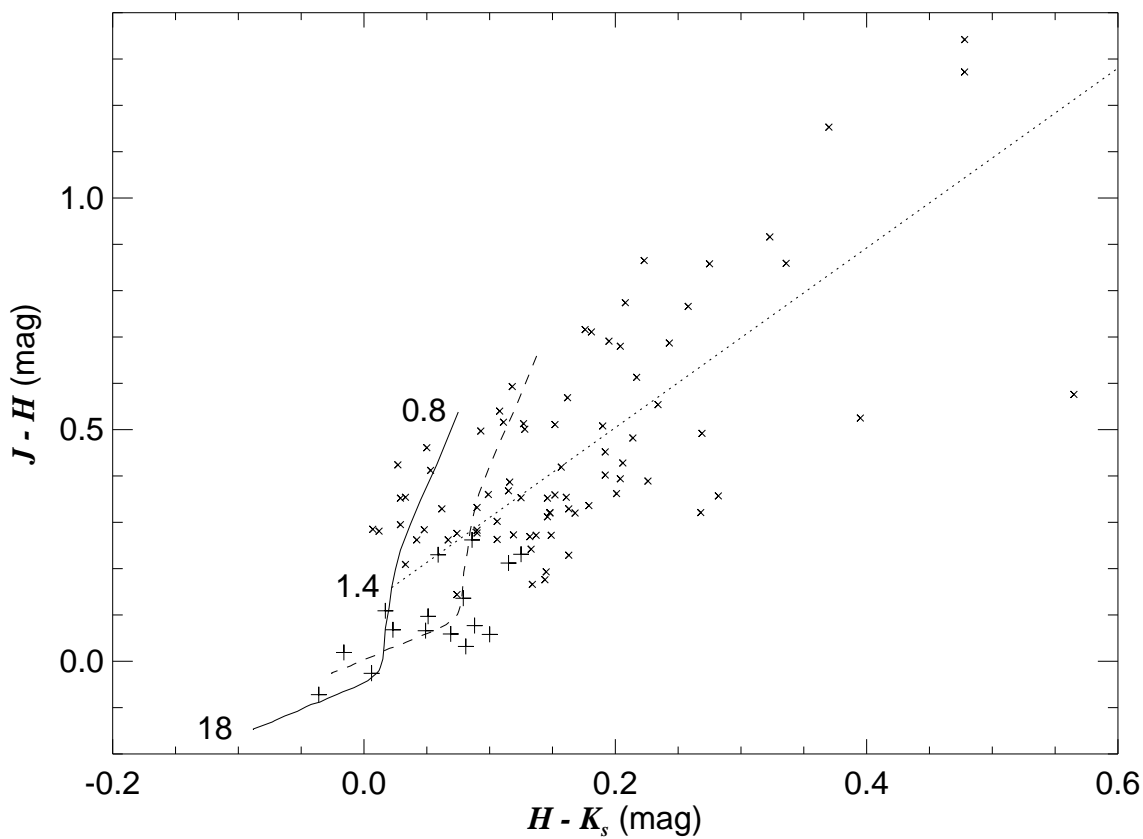}}
\end{center}
\caption[fig5]{The distribution of near-IR colors for those stars meeting the 
positional and proper motion criteria for cluster membership.  Plus signs mark 
those that are probable, main sequence members while $\times$ signs are used for
the rest of the sample.  The solid and dashed lines represent the values for 
the cluster isochrone for no reddening and $E(B-V)=0.37$, respectively. 
The numbers beside the unreddened isochrone indicate the stellar masses at 
the terminal points and at the origin of the reddening trajectory ({\it dotted line}).
Stars below the dotted line are candidate massive stars ($M> 1.4 M_\odot$).
\label{fig_ir_colors}} 
\end{figure}

We can apply a few other constraints to limit the number of 
field stars that may exist within the astrometric group 
of cluster candidates.  Figure \ref{fig_ir_colors} shows the near-IR color-color 
diagram for the sample together with unreddened ({\it solid line}) 
and reddened ({\it dashed line}) versions of the cluster isochrone.
The numbers beside the unreddened version indicate the stellar 
mass corresponding to the color pair, and massive stars have colors
in the lower part of the diagram.  The reddening vector 
direction is indicated by the dotted line.  Comer\'{o}n et al.\ (2002) and
others have argued that the reddening-free parameter $Q=(J-H)-1.94*(H-K_s)$ 
(Fitzpatrick 1999; Indebetouw et al.\ 2005) provides an excellent
means to identify massive, reddened stars that typically have $Q\approx 0.0$. 
We decided to apply this criterion by 
restricting the sample to those massive stars with $Q<0.116$, the model value
for a star of mass $M=1.4 M_\odot$.  Imposing this limit will mean that we lose fainter, 
red cluster members, but, on the other hand, we will also remove cool background giants and 
cooler, low mass, foreground stars that do not belong to the cluster
(for example, those stars found near the unreddened isochrone between 
0.8 and $1.4M_\odot$).   

\placefigure{fig5}     

We also applied a faint limit constraint, $R_u < 15$, to avoid stars
with large magnitude errors.  Since our initial sample was drawn from 
UCAC3 stars with complete photometric coverage, the result is that
this sample tends towards stars with equal magnitude in all filters 
at the faint end.  This group probably includes foreground A-type stars, 
so applying the faint limit will help us remove such stars from 
consideration. 
 
\begin{figure}
\begin{center}
{\includegraphics[angle=90,height=12cm]{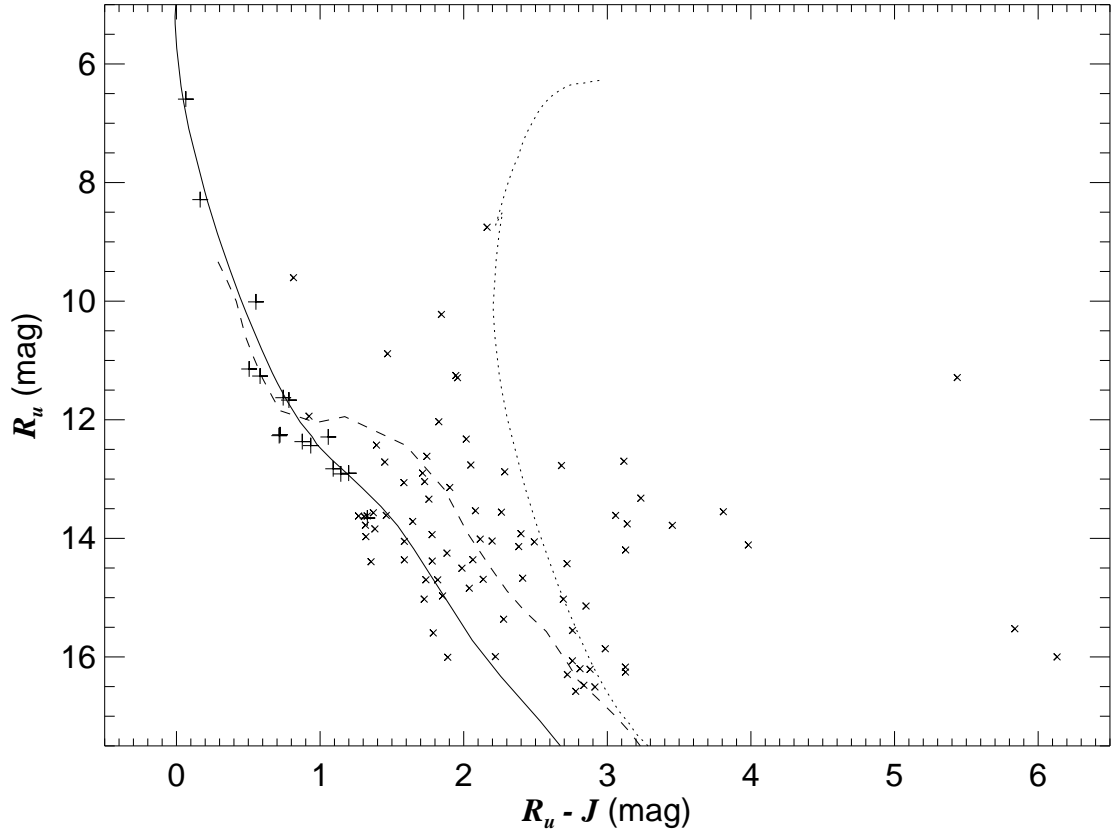}}
\end{center}
\caption[fig6]{A color-magnitude diagram based upon the UCAC3 $R_u$ and 2MASS $J$ 
magnitudes.  Plus signs mark probable, main sequence members, while  $\times$ 
symbols are used for the remainder of the sample that meets the positional and 
proper motion criteria for cluster membership.  The solid line shows the 
best fit isochrone. The dashed line indicates the pre-main sequence isochrone for an age of 7~Myr.  The dotted line is the isochrone for distant OB stars that may be related to the nearby cluster Berkeley~87.
\label{fig_ucac3_color_mag}}
\end{figure}

\begin{figure}
\begin{center}
{\includegraphics[angle=90,height=12cm]{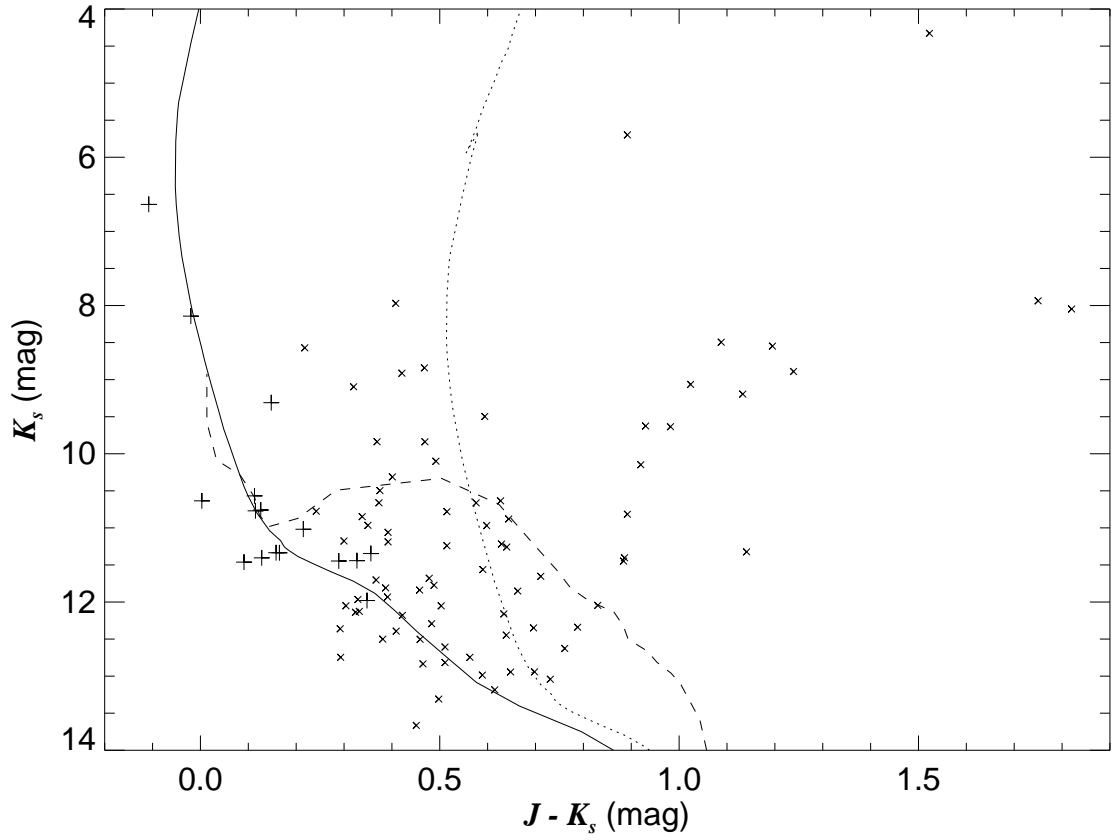}}
\end{center}
\caption[fig7]{A color-magnitude diagram for 2MASS $J$ and $K_s$  
magnitudes (in the same format as Fig.~\ref{fig_ucac3_color_mag}).  The bright, red object
({\it top, right}) is IRAS 20161+4035. 
\label{fig_2mass_color_mag}} 
\end{figure}

We show two versions of the color -- magnitude diagram in 
Figures \ref{fig_ucac3_color_mag} \& \ref{fig_2mass_color_mag}.  
Both of these show evidence of a main sequence 
group at the blue end, but both also show a large population 
of redder stars that lie above the expected main sequence. 
In order to remove such
targets, we restricted the sample to those with $R_u-J<1.35$. 
This constraint, together with the $Q$ limit described above, will 
result in a sample without any low mass, main sequence members. 
  
\placefigure{fig6}     

\placefigure{fig7}     

Our final constraint is based on how closely the magnitudes 
match the model isochrone values.  For any given choice of 
$E(B-V)$ and DM, we formed an average $\chi_\nu^2$ statistic
between all the observed and model magnitudes (reddened according 
to the relations given by Fitzpatrick 1999) for each mass point
along the model isochrone.  The best fit (lowest $\chi_\nu^2$)
model point was taken as an appropriate mass for the specific star.  
Those stars with magnitudes and colors very different from 
the main sequence values will have a large minimum $\chi_\nu^2$, 
and we restricted our final sample of main sequence cluster 
members to those with $\chi_\nu^2 < 35$, a value high enough 
to include those cases where the observational photometric
errors are underestimated.  The final sample consists of 
16 stars that are listed in Table~3 and are indicated in Figures 
\ref{fig_ir_colors} -- \ref{fig_2mass_color_mag} by plus signs.  
Note that we make the tacit assumption that the same reddening 
can be applied to all cluster members, and this can only be true
in some average sense.  There may be a reddening gradient 
across this field (Schlegel et al.\ 1998) and some of the youngest
objects may suffer from circumstellar dust reddening. 
Nevertheless, until spectroscopy is available for many stars and 
their intrinsic colors found, we must accept this working hypothesis.   

\placetable{tab3}      

We performed a grid search over a range of $E(B-V)$ and DM 
to find those values that minimized the average $\chi_\nu^2$
value for the selected cluster stars.  The best fit was made with 
$E(B-V)=0.37 \pm 0.05$ mag and DM $=9.35 \pm 0.03$ mag, 
yielding an average photometric statistic of $\chi_\nu^2 = 4.4$. 
The isochrones for these parameters are plotted in Figures \ref{fig_ir_colors} -- \ref{fig_2mass_color_mag}, 
and the corresponding main sequence masses from the isochrone 
are listed in the final column of Table 3.  
Our derived reddening estimate is mainly consistent with 
earlier estimates for the cluster and star 
($E(B-V)=0.34$, Kharchenko et al.\ 2005;
$E(B-V)=0.345$, Burnichon 1975;
$E(B-V)=0.38$, Schr\"{o}der et al.\ 2004;
$E(B-V)=0.41$, Cardelli et al.\ 1989).
The reddening is slightly larger than that found by SED fitting for the 
blue stars (Table~1), but we suspect that the differences 
arise in the variable reddening across the cluster. 
We also repeated the analysis using an isochrone for a 
younger cluster (age 3 Myr), but the results were unchanged 
because the predicted colors and magnitudes were only
significantly different for the most massive star 
HD~193322Aa1 (the predicted colors are marginally closer 
to the observed ones for the 7 Myr isochrone).

There are many stars that are plotted above the
main sequence in Figures~\ref{fig_ucac3_color_mag} 
 \& \ref{fig_2mass_color_mag}.  Some of these
stars may belong to a massive and more distant
population.  For example, in the infrared
color-color diagram (Fig.~\ref{fig_ir_colors}), there appears to
be a reddened group of OB stars near $H-K_s = 0.2$
and $J-H=0.4$.  An independent analysis of the 2MASS
magnitudes of stars in the vicinity of Cr~419
by David Turner (priv.\ communication) suggests that
these correspond to objects with $E(B-V)=1.5$ at
a distance of 1.6~kpc that may be associated with
the nearby cluster Berkeley~87 \citep{turner1982,massey2001}.
  We show how the 7~Myr isochrone
would appear for such reddened and distant stars
as a dotted line in Figures~\ref{fig_ucac3_color_mag} 
and \ref{fig_2mass_color_mag}, and many of the
stars appearing above the Cr~419 main sequence are
close to the predicted colors and magnitudes of
a distant population.  Cr~419 is probably young enough
that it may host pre-main sequence stars that would
also appear above the main sequence.  Dashed lines
in Figure~\ref{fig_ucac3_color_mag} 
and \ref{fig_2mass_color_mag} indicate the pre-main sequence
isochrone for an age of 7~Myr from the work of \citet{siess2000}
 transformed to the distance
and reddening of Cr~419, and indeed we find many
examples of stars close to the predicted track.
We suspect that redder stars appearing in these
color-magnitude diagrams probably include some
cluster pre-main sequence stars and large numbers of
distant massive stars and other line-of-sight field stars.


\section{$UBV$ Photometry}  

We also obtained new $UBV$ photometry of the central region of 
the cluster to explore the optical colors of the fainter stars.
One of us (JRP) collected CCD images of Collinder 419 in 2006 October 
with the Emory University Observatory, DFM 0.6~m Cassegrain 
telescope\footnote{http://www.physics.emory.edu/astronomy/observatory.html}.  
The detector was an Apogee Ap47 $1024\times1024$ pixel CCD camera.  
In order to concentrate the signal for fainter sources, the camera was set 
to a $2\times2$ binning mode with a net pixel scale of $1\farcs1$ pixel$^{-1}$ 
and a $9\farcm5 \times 9\farcm5$ field of view.  The images of the brightest 
stars were saturated in individual frames in order to obtain a stronger signal 
on the faintest sources.  Thirty frames were collected through each Johnson $UBV$ 
filter.  In addition, standard stars from charts 140, 141, and 148 of Landolt (1992) were observed over a large range in air mass to transform the instrumental
magnitudes to the standard system.  

All the frames were debiased, dark subtracted, and flat fielded using standard 
routines in IRAF.  The images were combined using IMCENTROID and IMCOMBINE 
to build a master image in each filter.  Photometric measurements of the standard stars were 
made with aperture photometry routines in IRAF.  The transformation 
coefficients were computed using the IDL procedure transf1.pro, written by Marc W.\ Buie.  
This procedure uses the general formula
$$m_O = m_i - k' X - k'' C X + e C + Z$$
where $m_0$ is the standard magnitude, $m_i$ is the instrumental magnitude, 
$k'$ is the first order extinction coefficient, $X$ is the airmass, $k''$ is the 
second order extinction coefficient, $C$ is a color index,
$e$ is a color coefficient, and $Z$ is the magnitude zero point. 
The instrumental magnitudes of the stars in the Collinder 419 frames
were measured using PSF fitting routines in IRAF, and the relation between the 
PSF and aperture measurements was determined for ten well isolated stars in the field.  
The transformation equations were then used to determine Johnson magnitudes for 
the cluster stars.  The errors in the transformation coefficients were added in 
quadrature with the instrumental magnitude errors to obtain the net observational errors
(approximately 0.04, 0.03, and 0.02 mag in $UBV$, respectively).  The results
are listed in Table 4. 

\placetable{tab4}      

\begin{figure}
\begin{center}
{\includegraphics[angle=90,height=12cm]{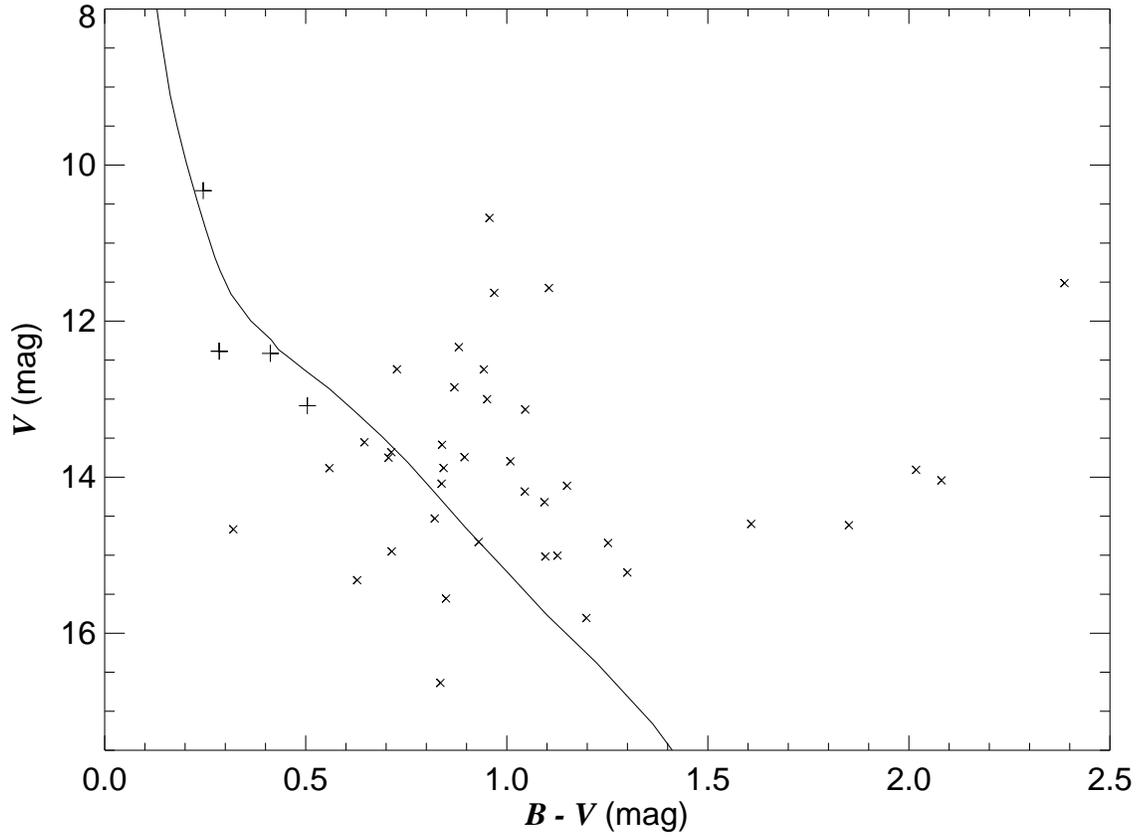}}
\end{center}
\caption[fig8]{A color-magnitude diagram from optical photometry in 
the same format as Fig.~~\ref{fig_ucac3_color_mag}.  The solid line shows the isochrone for our derived $E(B-V)$ and DM.
\label{fig_optical_color_mag}} 
\end{figure}

\begin{figure}
\begin{center}
{\includegraphics[angle=90,height=12cm]{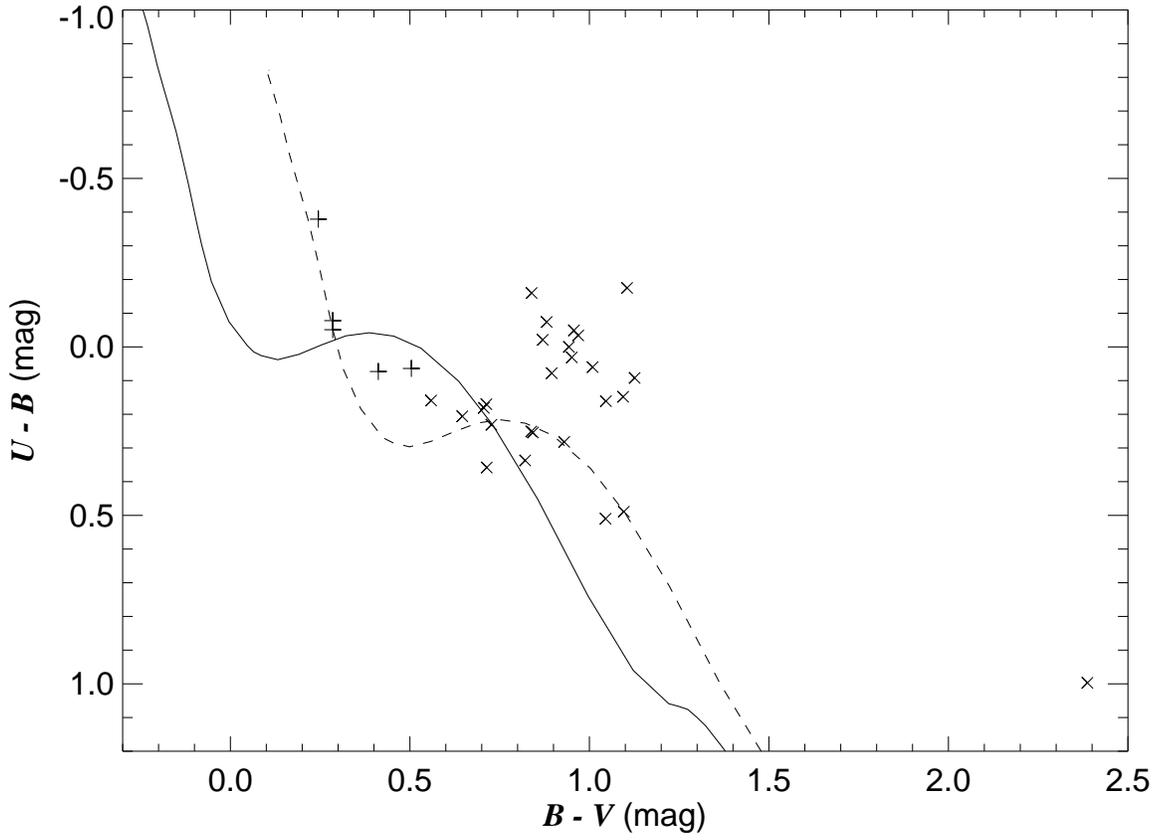}}
\end{center}
\caption[fig9]{A color-color diagram from optical photometry 
in a format similar to Fig.~\ref{fig_ir_colors}.  The solid and dashed lines show the 
model unreddened and reddened colors, respectively.
\label{fig_color_color}} 
\end{figure}

We show in Figure \ref{fig_optical_color_mag} the color -- magnitude diagram $(B-V,V)$ for 43 of the 79 stars
measured that meet the proper motion limits for cluster membership described above (\S3). 
Most of the proposed cluster members (\S3; Table~3) were too bright and/or outside of
the field of view for our $UBV$ photometric measurements, and we have $UBV$ 
photometry for only 5 of the 16 proposed members (indicated by plus signs in 
Fig.~8 and by ``Yes'' in the final column of Table~4).  Figure \ref{fig_optical_color_mag} also shows the 
main sequence relation from the models of Bertelli et al.\ (1994) and 
Marigo et al.\ (2008) for our derived estimates of $E(B-V)=0.37$ mag and 
DM = 9.35 mag, and this relation appears to be consistent with our 
optical photometry.  There are a number of stars that have magnitudes 
indicative of lower mass, main sequence stars, but there are many objects
with a position above the main sequence with a distribution similar to 
that seen in the red and near-infrared color -- magnitude diagrams (Figs. \ref{fig_ucac3_color_mag} and \ref{fig_2mass_color_mag}).
Figure \ref{fig_color_color} shows the optical color - color diagram $(B-V, U-B)$ for the 
subset with complete $UBV$ measurements and within the proper motion limits 
for membership.  We also show both unreddened and reddened versions of the 
model colors.  Again, there is evidence for some lower main sequence members
(towards the middle and lower sections of the dashed line), but there is also 
a group of stars with colors near $(B-V, U-B) \approx (1.0, 0.0)$ that are 
characteristic of B-stars with a reddening $E(B-V)\approx 1.0$ mag.  Most of 
these stars also have positions in the near-IR color -- color diagram 
associated with reddened, intermediate mass stars (Fig. \ref{fig_ir_colors};
$(H-K_s, J-H) \approx (0.15, 0.30)$).  

\placefigure{fig8}     

\placefigure{fig9}     


\section{Discussion}  

The distance we obtained from the photometric analysis of 
$d = 741 \pm 11$ pc (formal error) does not account for 
possible systematic errors in the model isochrone magnitudes. 
By comparing the $(R_c-J,J)$ color-magnitude diagrams for 
a 7 Myr cluster from Marigo et al.\ (2008) with that 
from Lejeune \& Schaerer (2001), we estimate that 
the model uncertainties amount to $\triangle J \approx 0.10$ mag over the 
main sequence range of interest, so we add this error
in quadrature with the fitting errors to obtain $d = 741 \pm 36$ pc.
We can check this with distances 
derived from the angular sizes given in Table~1.  
This distance estimate depends on the ratio of the physical diameter 
to the angular diameter, and for this comparison, we 
estimated the stellar radii for HD~193322B, C, and D 
from the model isochrone values for main sequence stars
with our adopted effective temperatures (Table~2). 
We could not use this relation for HD~193322Aa1 because
the model isochrone did not extend to stars quite this hot, 
so we instead adopted a radius of $R=7.53 R_\odot$ from 
the spectral calibration for main sequence stars from 
Martins et al.\ (2005).  We estimated a radius of 
$R=75 R_\odot$ for the M3~III star IRAS 20161+4035 from 
the calibration for giants by van Belle et al.\ (1999). 
The resulting distances are listed in the final column 
of Table~1, and the average, $d = 766 \pm 51$ pc, agrees 
with the photometric result.  Our results confirm 
the only previous distance estimate for the cluster from 
Kharchenko et al.\ (2005) of 740 pc (no error quoted). 
The cluster distance also agrees within errors with the
{\it Hipparcos} distance for HD~193322 of $600^{+150}_{-110}$ pc
(van Leeuwen 2007).

We suspect that some of the red stars that meet the spatial 
and proper motion constraints of cluster membership but lie 
above the main sequence are very young objects.  A similar 
population of bright, red objects has been reported for
other young star clusters containing O stars 
(Kumar et al.\ 2004;  Comer\'{o}n \& Pasquali 2005;
Carlson et al.\ 2007; Negueuela et al.\ 2007; Comer\'{o}n et al.\ 2008).
These stars may include IR-excess objects with remnant disks like
the Herbig Ae/Be stars.  Although it is possible that the
red object, IRAS 20161+4035, is an old star that is coincidentally
passing through the vicinity of Cr~419, we suspect that it
may also be a very young member of the cluster, despite its 
evolved appearance.   The star's large reddening and IR-excess 
both suggest that it is immersed in natal dust, and its
strong \ion{Li}{1} $\lambda 6707$ and high luminosity may be 
indicative of youth.  Kumar et al.\ (2004) identified three 
other cases of luminous, late-type giants that are found 
in young clusters.   These cool giants may represent another 
kind of spectroscopic manifestation of young stellar objects.


\acknowledgments

We thank Todd Henry, Adric Riedel, and Russel White for helpful comments about M-star spectra, and we also thank Bill Binkert of the KPNO staff for his assistance at the KPNO Coude Feed Telescope. We are grateful to David Turner for discussions about
field contamination by distant massive stars.  This publication made use of data products from the Two Micron All Sky  Survey, which is a joint project of the University of Massachusetts and  the Infrared Processing and Analysis Center/California Institute of Technology, funded by the National Aeronautics and Space Administration and the National Science Foundation.  It also used images from the Second Palomar Observatory Sky Survey (POSS-II), which was made by the California Institute of Technology with funds from the National Science Foundation, the National Geographic Society, the Sloan Foundation, the Samuel Oschin Foundation, and the Eastman Kodak Corporation. We also made use of the Washington Double Star Catalog, maintained at the U.\ S.\ Naval Observatory, and the SIMBAD database, operated by the CDS in Strasbourg, France. This material is based upon work supported by the National Science Foundation  under Grant No.~AST-0606861. Institutional support has been provided from the GSU College of Arts and Sciences and from the Research Program Enhancement fund of the Board of Regents of the University System of Georgia, administered through the GSU Office of the Vice President for Research.  A portion of the research  was carried out at the Jet Propulsion Laboratory, California Institute of Technology, under a contract with the National Aeronautics and Space Administration.



\clearpage

 

\clearpage



\begin{deluxetable}{lcccc}
\tablenum{1}
\tablewidth{0pt}
\tabletypesize{\scriptsize}
\tablecaption{Reddening Estimates for $R=3.1$\label{tab1}} 
\tablehead{
\colhead{Star} & 
\colhead{Spectral} & 
\colhead{$E(B-V)$} &
\colhead{$\theta_{LD}$} &
\colhead{$d$}
\\ 
\colhead{Name}  & 
\colhead{Classification}  & 
\colhead{(mag)}  &
\colhead{($\mu$as)}  &
\colhead{(kpc)}
} 
\startdata 
HD 193322Aa1\dotfill    & O9 V:((n))  & $0.31\pm0.02$ &  $85\pm2$  & 0.83 \\ 
HD 193322B\dotfill      & B1.5 V      & $0.28\pm0.03$ &  $53\pm2$  & 0.73 \\ 
HD 193322C\dotfill      & B8 V        & $0.29\pm0.02$ &  $23\pm1$  & 0.79 \\
HD 193322D\dotfill      & B9 V        & $0.29\pm0.03$ &  $25\pm1$  & 0.70 \\ 
IRAS 20161+4035\dotfill & M3 III      & $0.74\pm0.04$ & $893\pm46$ & 0.78 \\
\enddata
\end{deluxetable}


\begin{deluxetable}{lcccccccc}
\tablenum{2}
\tablewidth{0pt}
\tabletypesize{\scriptsize}
\tablecaption{Adopted Flux Ratios for the Components of HD 193322\label{tab2}} 
\tablehead{
\colhead{Star} & 
\colhead{$T_{\rm eff}$} & 
\colhead{$F_x/F_{Aa1}$} &
\colhead{$F_x/F_{Aa1}$} &
\colhead{$F_x/F_{Aa1}$} &
\colhead{$F_x/F_{Aa1}$} &
\colhead{$F_x/F_{Aa1}$} &
\colhead{$F_x/F_{Aa1}$} &
\colhead{$F_x/F_{Aa1}$} 
\\ 
\colhead{Name}  & 
\colhead{(kK)}  & 
\colhead{$B$}   &
\colhead{$V$}   &
\colhead{$R_j$} &
\colhead{$I_j$} &
\colhead{$J$}   &
\colhead{$H$}   &
\colhead{$K_s$} 
} 
\startdata 
HD 193322Aa1\dotfill    & 33.2 &  1.00 &  1.00 &  1.00 &  1.00 &  1.00 &  1.00 &  1.00 \\ 
HD 193322Aa2\dotfill    & 23.0 &  0.29 &  0.30 &  0.31 &  0.33 &  0.33 &  0.35 &  0.35 \\ 
HD 193322Ab\dotfill     & 25.0 &  0.41 &  0.43 &  0.44 &  0.46 &  0.47 &  0.48 &  0.48 \\ 
HD 193322B\dotfill      & 23.0 &  0.21 &  0.21 &  0.21 &  0.21 &  0.23 &  0.24 &  0.25 \\ 
\enddata
\end{deluxetable}


\begin{deluxetable}{cccccccc}
\tablenum{3}
\tablewidth{0pt}
\tabletypesize{\scriptsize}
\tablecaption{Main Sequence Cluster Members\label{tab3}} 
\tablehead{
\colhead{UCAC3} & 
\colhead{Other} & 
\colhead{$R_u$} & 
\colhead{$R_u - J$} & 
\colhead{$J - H$} & 
\colhead{$H - K_s$} & 
\colhead{$K_s$} & 
\colhead{Mass} 
\\ 
\colhead{Number} & 
\colhead{Name} & 
\colhead{(mag)}  & 
\colhead{(mag)}  & 
\colhead{(mag)}  & 
\colhead{(mag)}  & 
\colhead{(mag)}  & 
\colhead{($M_\odot$)}  
} 
\startdata
262$-$202014 & \nodata      &     12.29 &  1.06 & \phs 0.14 & \phs 0.08 &     11.02 & \phn 2.2 \\
262$-$202033 & \nodata      &     12.90 &  1.20 & \phs 0.23 & \phs 0.12 &     11.35 & \phn 1.7 \\
262$-$202070 & HD 228810    &     10.01 &  0.55 & \phs 0.10 & \phs 0.05 & \phn 9.31 & \phn 5.0 \\
262$-$202090 & \nodata      &     12.44 &  0.94 & \phs 0.08 & \phs 0.09 &     11.34 & \phn 1.9 \\
262$-$202099 & \nodata      &     13.66 &  1.33 & \phs 0.26 & \phs 0.09 &     11.98 & \phn 1.4 \\
262$-$202103 & \nodata      &     12.91 &  1.14 & \phs 0.21 & \phs 0.12 &     11.44 & \phn 1.7 \\
262$-$202123 & HD 193322C   &     11.14 &  0.51 & \phs 0.02 &   $-$0.02 &     10.63 & \phn 2.8 \\
262$-$202132 & HD 193322D   &     11.26 &  0.58 & \phs 0.03 & \phs 0.08 &     10.57 & \phn 2.8 \\
262$-$202133 & HD 193322Aa1 & \phn 6.59 &  0.07 &   $-$0.07 &   $-$0.04 & \phn 6.64 &     14.9 \\
262$-$202133 & HD 193322B   & \phn 8.29 &  0.17 &   $-$0.03 & \phs 0.01 & \phn 8.14 & \phn 9.0 \\
262$-$202155 & \nodata      &     12.27 &  0.71 & \phs 0.07 & \phs 0.02 &     11.46 & \phn 1.8 \\
262$-$202269 & \nodata      &     12.25 &  0.72 & \phs 0.06 & \phs 0.07 &     11.40 & \phn 1.8 \\
262$-$202314 & \nodata      &     12.83 &  1.09 & \phs 0.23 & \phs 0.06 &     11.45 & \phn 1.7 \\
262$-$202350 & \nodata      &     12.37 &  0.88 & \phs 0.06 & \phs 0.10 &     11.34 & \phn 1.9 \\
262$-$202365 & \nodata      &     11.63 &  0.74 & \phs 0.07 & \phs 0.05 &     10.77 & \phn 2.5 \\
262$-$202402 & \nodata      &     11.67 &  0.78 & \phs 0.11 & \phs 0.02 &     10.76 & \phn 2.5 \\
\enddata
\end{deluxetable}


\begin{deluxetable}{ccccc}
\tablenum{4}
\tablewidth{0pt}
\tabletypesize{\scriptsize}
\tablecaption{$UBV$ Photometry\label{tab4}} 
\tablehead{
\colhead{UCAC3} & 
\colhead{$U$} & 
\colhead{$B$} & 
\colhead{$V$} & 
\colhead{} 
\\ 
\colhead{Number} & 
\colhead{(mag)}  & 
\colhead{(mag)}  & 
\colhead{(mag)}  & 
\colhead{Member}  
} 
\startdata
  262-201944 & 12.51$\pm$0.04 & 12.68$\pm$0.02 & 11.58$\pm$0.01 &\nodata \\
  262-201945 & 14.26$\pm$0.04 & 14.42$\pm$0.02 & 13.59$\pm$0.02 &\nodata \\
  262-201948 &  \nodata       & 15.92$\pm$0.03 & 13.91$\pm$0.02 &\nodata \\
  262-201953 &  \nodata       & 16.52$\pm$0.04 & 15.21$\pm$0.01 &\nodata \\
  262-201963 & 16.25$\pm$0.05 & 16.09$\pm$0.03 & 14.91$\pm$0.02 &\nodata \\
  262-201973 & 16.22$\pm$0.06 & 16.13$\pm$0.04 & 15.01$\pm$0.02 &\nodata \\
  262-201978 & 16.60$\pm$0.06 & 16.11$\pm$0.04 & 15.02$\pm$0.02 &\nodata \\
  262-201981 & 15.30$\pm$0.04 & 15.27$\pm$0.02 & 14.43$\pm$0.01 &\nodata \\
  262-201993 & 16.04$\pm$0.05 & 15.76$\pm$0.02 & 14.83$\pm$0.02 &\nodata \\
  262-201997 &  \nodata       & 16.66$\pm$0.03 & 15.53$\pm$0.03 &\nodata \\
  262-202009 &  \nodata       & 17.15$\pm$0.05 & 15.86$\pm$0.01 &\nodata \\
  262-202014 & 12.90$\pm$0.04 & 12.83$\pm$0.01 & 12.41$\pm$0.03 & Yes    \\
  262-202017 & 15.74$\pm$0.04 & 15.23$\pm$0.02 & 14.18$\pm$0.01 &\nodata \\
  262-202018 &  \nodata       & 14.99$\pm$0.04 & 14.67$\pm$0.03 &\nodata \\
  262-202025 &  \nodata       & 16.47$\pm$0.03 & 14.62$\pm$0.01 &\nodata \\
  262-202026 &  \nodata       & 16.31$\pm$0.03 & 15.22$\pm$0.01 &\nodata \\
  262-202027 &  \nodata       & 16.49$\pm$0.03 & 15.61$\pm$0.01 &\nodata \\
  262-202032 & 14.72$\pm$0.04 & 14.64$\pm$0.01 & 13.74$\pm$0.01 &\nodata \\
  262-202033 & 13.65$\pm$0.04 & 13.59$\pm$0.01 & 13.08$\pm$0.02 & Yes    \\
  262-202034 &  \nodata       & 16.12$\pm$0.02 & 14.04$\pm$0.02 &\nodata \\
  262-202039 & 15.17$\pm$0.04 & 14.92$\pm$0.01 & 14.08$\pm$0.01 &\nodata \\
  262-202042 &  \nodata       & 15.95$\pm$0.02 & 15.32$\pm$0.02 &\nodata \\
  262-202045 & 14.40$\pm$0.04 & 14.20$\pm$0.03 & 13.55$\pm$0.03 &\nodata \\
  262-202046 &  \nodata       & 16.88$\pm$0.04 & 15.73$\pm$0.01 &\nodata \\
  262-202047 &  \nodata       & 17.26$\pm$0.05 & 15.96$\pm$0.01 &\nodata \\
  262-202050 &  \nodata       & 16.52$\pm$0.03 & 15.22$\pm$0.01 &\nodata \\
  262-202052 & 15.33$\pm$0.04 & 14.99$\pm$0.02 & 14.20$\pm$0.03 &\nodata \\
  262-202054 & 16.21$\pm$0.05 & 15.74$\pm$0.02 & 14.76$\pm$0.02 &\nodata \\
  262-202059 & 14.90$\pm$0.04 & 13.90$\pm$0.01 & 11.51$\pm$0.04 &\nodata \\
  262-202061 &  \nodata       & 17.47$\pm$0.05 & 16.64$\pm$0.01 &\nodata \\
  262-202066 & 14.98$\pm$0.04 & 14.73$\pm$0.02 & 13.88$\pm$0.03 &\nodata \\
  262-202067 & 14.64$\pm$0.04 & 14.46$\pm$0.01 & 13.75$\pm$0.02 &\nodata \\
  262-202070 & 10.19$\pm$0.04 & 10.57$\pm$0.01 & 10.33$\pm$0.01 & Yes    \\
  262-202088 & 13.09$\pm$0.04 & 13.03$\pm$0.01 & 12.48$\pm$0.02 &\nodata \\
  262-202089 & 16.76$\pm$0.07 & 16.24$\pm$0.03 & 15.13$\pm$0.02 &\nodata \\
  262-202097 &  \nodata       & 16.21$\pm$0.03 & 14.60$\pm$0.01 &\nodata \\
  262-202098 & 15.69$\pm$0.05 & 15.35$\pm$0.02 & 14.53$\pm$0.01 &\nodata \\
  262-202100 &  \nodata       & 17.24$\pm$0.05 & 15.92$\pm$0.02 &\nodata \\
  262-202107 & 16.02$\pm$0.05 & 15.67$\pm$0.02 & 14.95$\pm$0.02 &\nodata \\
  262-202109 &  \nodata       & 16.33$\pm$0.03 & 15.31$\pm$0.02 &\nodata \\
  262-202117 & 13.98$\pm$0.04 & 13.95$\pm$0.01 & 13.00$\pm$0.02 &\nodata \\
  262-202118 &  \nodata       & 16.08$\pm$0.05 & 15.53$\pm$0.01 &\nodata \\
  262-202120 &  \nodata       & 15.26$\pm$0.02 & 14.11$\pm$0.02 &\nodata \\
  262-202128 & 13.56$\pm$0.04 & 13.56$\pm$0.01 & 12.62$\pm$0.01 &\nodata \\
  262-202135 & 16.41$\pm$0.05 & 16.15$\pm$0.02 & 15.12$\pm$0.01 &\nodata \\
  262-202141 & 16.66$\pm$0.09 & 15.72$\pm$0.03 & 14.48$\pm$0.01 &\nodata \\
  262-202146 &  \nodata       & 16.35$\pm$0.03 & 15.13$\pm$0.04 &\nodata \\
  262-202155 & 12.62$\pm$0.04 & 12.67$\pm$0.01 & 12.39$\pm$0.03 & Yes    \\
  262-202156 &  \nodata       & 16.79$\pm$0.03 & 15.72$\pm$0.02 &\nodata \\
  262-202159 & 16.45$\pm$0.05 & 16.09$\pm$0.02 & 15.08$\pm$0.01 &\nodata \\
  262-202161 &  \nodata       & 17.00$\pm$0.03 & 16.07$\pm$0.02 &\nodata \\
  262-202166 & 14.60$\pm$0.04 & 14.44$\pm$0.02 & 13.88$\pm$0.02 &\nodata \\
  262-202170 & 11.59$\pm$0.04 & 11.63$\pm$0.01 & 10.68$\pm$0.03 &\nodata \\
  262-202175 &  \nodata       & 17.59$\pm$0.08 & 16.33$\pm$0.05 &\nodata \\
  262-202176 & 12.57$\pm$0.04 & 12.61$\pm$0.01 & 11.64$\pm$0.01 &\nodata \\
  262-202177 & 15.22$\pm$0.04 & 14.70$\pm$0.02 & 13.65$\pm$0.04 &\nodata \\
  262-202178 &  \nodata       & 16.73$\pm$0.04 & 15.81$\pm$0.01 &\nodata \\
  262-202185 &  \nodata       & 17.45$\pm$0.06 & 16.09$\pm$0.01 &\nodata \\
  262-202187 &  \nodata       & 17.41$\pm$0.06 & 16.30$\pm$0.01 &\nodata \\
  262-202190 & 14.87$\pm$0.04 & 14.81$\pm$0.01 & 13.80$\pm$0.01 &\nodata \\
  262-202194 & 15.97$\pm$0.05 & 15.68$\pm$0.02 & 14.71$\pm$0.03 &\nodata \\
  262-202197 &  \nodata       & 16.89$\pm$0.04 & 15.79$\pm$0.01 &\nodata \\
  262-202199 & 14.34$\pm$0.04 & 14.18$\pm$0.01 & 13.13$\pm$0.01 &\nodata \\
  262-202207 & 14.56$\pm$0.04 & 14.39$\pm$0.01 & 13.68$\pm$0.01 &\nodata \\
  262-202212 & 15.56$\pm$0.04 & 15.41$\pm$0.02 & 14.32$\pm$0.02 &\nodata \\
  262-202218 & 13.14$\pm$0.04 & 13.21$\pm$0.01 & 12.33$\pm$0.01 &\nodata \\
  262-202225 &  \nodata       & 16.96$\pm$0.04 & 15.83$\pm$0.02 &\nodata \\
  262-202224 &  \nodata       & 16.75$\pm$0.03 & 15.66$\pm$0.01 &\nodata \\
  262-202227 & 16.31$\pm$0.05 & 15.84$\pm$0.02 & 14.89$\pm$0.06 &\nodata \\
  262-202236 & 16.47$\pm$0.05 & 16.18$\pm$0.03 & 15.41$\pm$0.01 &\nodata \\
  262-202239 &  \nodata       & 17.00$\pm$0.04 & 15.81$\pm$0.04 &\nodata \\
  262-202243 & 16.16$\pm$0.05 & 15.65$\pm$0.02 & 14.80$\pm$0.01 &\nodata \\
  262-202247 & 13.70$\pm$0.04 & 13.72$\pm$0.01 & 12.85$\pm$0.03 &\nodata \\
  262-202248 & 13.57$\pm$0.04 & 13.34$\pm$0.02 & 12.62$\pm$0.07 &\nodata \\
  262-202255 &  \nodata       & 16.40$\pm$0.03 & 15.56$\pm$0.01 &\nodata \\
  262-202256 &  \nodata       & 17.86$\pm$0.10 & 16.15$\pm$0.03 &\nodata \\
  262-202260 &  \nodata       & 16.10$\pm$0.03 & 14.84$\pm$0.01 &\nodata \\
  262-202265 & 14.56$\pm$0.04 & 14.08$\pm$0.02 & 13.27$\pm$0.01 &\nodata \\
  262-202269 & 12.59$\pm$0.04 & 12.67$\pm$0.02 & 12.38$\pm$0.02 & Yes    \\
\enddata
\end{deluxetable}


\clearpage

\begin{thebibliography}{}
\bibitem[Abt \& Cardona(1983)]{abt83}
         Abt, H. A., \& Cardona, O. 1983, \apj, 272, 182
\bibitem[Barnard(1927)]{bar27} 
         Barnard, E. E. 1927, A photographic atlas of selected regions of the Milky Way,
         ed. E. B. Frost \& M. R. Calvert (Carnegie Inst. Washington Publ. 247) 
         (Washington DC: Carnegie Inst.)
\bibitem[Bertelli et al.(1994)]{ber94}
         Bertelli, G., Bressan, A., Chiosi, C., Fagotto, F., \& Nasi, E.
         1994, \aaps, 106, 275
\bibitem[Bessell et al.(1998)]{bes98}
         Bessell, M. S., Castelli, F., \& Plez, B. 1998, \aap, 333, 231
\bibitem[B\"{o}hm-Vitense(1981)]{boh81}
         B\"{o}hm-Vitense, E. 1981, \araa, 19, 295
\bibitem[Burnichon(1975)]{bur75}
         Burnichon, M. L. 1975, \aap, 45, 383
\bibitem[Burnichon \& Garnier(1976)]{bur76}
         Burnichon, M. L., \& Garnier, R. 1976, \aaps, 24, 89
\bibitem[Cardelli et al.(1989)]{car89}
         Cardelli, J. A., Clayton, G. C., \& Mathis, J. S. 1989, \apj, 345, 245
\bibitem[Carlson et al.(2007)]{car07}
         Carlson, L. R., et al. 2007, \apj, 665, L109
\bibitem[Cohen et al.(2003)Cohen, Wheaton, \& Megeath]{coh03}
         Cohen, M., Wheaton, W. A., \& Megeath, S. T.	
         2003, \aj, 126, 1090
\bibitem[Colina et al.(1996)Colina, Bohlin, \& Castelli]{col96}
         Colina, L., Bohlin, R., \& Castelli, F. 1996, HST Instrument
         Science Report CAL/SCS-008 (Baltimore: STScI)
\bibitem[Collinder(1931)]{col31}
         Collinder, P. 1931,  Ann. Obs. Lund, 2, 1
\bibitem[Comer\'{o}n \& Pasquali(2005)]{com05}
         Comer\'{o}n, F., \& Pasquali, A. 2005, \aap, 430, 541
\bibitem[Comer\'{o}n et al.(2008)]{com08}
         Comer\'{o}n, F., Pasquali, A., Figueras, F., \& Torra, J. 2008, \aap, 486, 453
\bibitem[Comer\'{o}n et al.(2002)]{com02}
         Comer\'{o}n, F., et al. 2002, \aap, 389, 874 
\bibitem[Cutri et al.(2003)]{cut03}
         Cutri, R., et al.\ 2003, 2MASS All-Sky Catalog of Point Sources (Pasadena: IPAC/Caltech)
\bibitem[Droege et al.(2006)]{dro06}
         Droege, T. F., Richmond, M. W., \& Sallman, M. 2006, \pasp, 118, 1666
\bibitem[Eaton(1995)]{eat95}
         Eaton, J. A. 1995, \aj, 109, 1797
\bibitem[Egan \& Price(1996)]{ega96}
         Egan, M. P., \& Price, S. D. 1996, \aj, 112, 2862
\bibitem[Fitzpatrick(1999)]{fit99}
         Fitzpatrick, E. L. 1999, \pasp, 111, 63
\bibitem[Fullerton(1990)]{ful90} 
         Fullerton, A. W. 1990, Ph.D. dissertation, Univ. of Toronto
\bibitem[Gustafsson et al.(2008)]{gus08}
         Gustafsson, B., Edvardsson, B., Eriksson, K., J\o rgensen, U. G., Nordlund, A., 
         \& Plez, B. 2008, \aap, 486, 951
\bibitem[Hartkopf et al.(1993)]{har93}
         Hartkopf, W. I., Gies, D. R., Mason, B. D., Bagnuolo, W. G., Jr., \& 
         McAlister, H. A. 1993, BAAS, 25, 872
\bibitem[Hoffleit(1982)]{hof82}
         Hoffleit, D. 1982, The Bright Star Catalogue, 4th ed. (New Haven: Yale Univ. Obs.) 
\bibitem[Indebetouw et al.(2005)]{ind05}
         Indebetouw, R., et al. 2005, \apj, 619, 931
\bibitem[Kharchenko et al.(2005)]{kha05}
         Kharchenko, N. V., Piskunov, A. E., R\"{o}ser, S., Schilbach, E., \& Scholz, R.-D.
         2005, \aap, 438, 1163         
\bibitem[Kumar et al.(2004)]{kum04}
         Kumar, B., Sagar, R., Sanwal, B. B., \& Bessell, M. S. 2004, \mnras, 353, 991
\bibitem[Landolt(1992)]{lan92}
         Landolt, A. U. 1992, \aj, 104, 340
\bibitem[Lejeune \& Schaerer(2001)]{lej01}
         Lejeune, T., \& Schaerer, D. 2001, \aap, 366, 538
\bibitem[Lesh(1979)]{les79}
         Lesh, J. R. 1979, in Spectral Classification of the Future (IAU Coll.\ 47),
         Ricerche Astronomiche, 9, ed. M. F. McCarthy, A. D. G. Philip, \& G. V. Coyne
         (Citta del Vaticano: Specola Vaticana), 81 
\bibitem[Marigo et al.(2008)]{mar08}
         Marigo, P., Girardi, L., Bressan, A., Groenewegen, M. A. T., Silva, L., \& Granato, G. L.
         2008, \aap, 482, 883
\bibitem[Martins et al.(2005)]{mar05}
         Martins, F., Schaerer, D., \& Hillier, D. J. 2005, \aap, 436, 1049
\bibitem[Mason et al.(2009)]{mas09}
         Mason, B. D., Hartkopf, W. I., Gies, D. R., Henry, T. J., \& Helsel, J. W.
         2009, AJ, 137, 3358
\bibitem[Massey et al.(2001)]{massey2001}
         Massey, P., DeGioia-Eastwood, K., \& Waterhouse, E. 2001, \aj, 121, 1050
\bibitem[McAlister et al.(1987)]{mca87}
         McAlister, H. A., Hartkopf, W. I., Hutter, D. J., Shara, M. M., \& Franz, O. G. 1987, \aj, 92, 183
\bibitem[McKibben et al.(1998)]{mck98}
         McKibben, W. P., et al. 1998, \pasp, 110, 900
\bibitem[Negueruela et al.(2007)]{neg07}
         Negueruela, I., Marco, A., Israel, G. L., \& Bernabeu, G.
         2007, \aap, 471, 485
\bibitem[Perryman et al.(1997)]{per97}
         Perryman, M. A. C., et al.  1997, \aap, 323, L49 
\bibitem[Reid et al.(1991)]{reid1991}
         Reid, I.N., et al. 1991, \pasp, 103, 661
\bibitem[Schilbach \& R\"{o}ser(2008)]{sch08}
         Schilbach, E., \& R\"{o}ser, S. 2008, \aap, 489, 105
\bibitem[Schlegel et al.(1998)]{sch98}
         Schlegel, D. J., Finkbeiner, D. P., \& Davis, M. 1998, \apj, 500, 525
\bibitem[Schr\"{o}der et al.(2004)]{sch04}
         Schr\"{o}der, S. E., Kaper, L., Lamers, H. J. G. L. M.,
         \& Brown, A. G. A. 2004, \aap, 428, 149
\bibitem[Siess et al.(2000)]{siess2000}
         Siess, L., Dufour, E., \& Forestini, M. 2000, \aap, 358, 593
\bibitem[Shatskii \& Tokovinin(1998)]{sha98}
         Shatskii, N. I., \& Tokovinin, A. A. 1998, Astronomy Letters, 24, 673
\bibitem[ten Brummelaar et al.(2000)]{tbr00} 
         ten Brummelaar, T., Mason, B. D., McAlister, H. A., Roberts, L. C., Jr., 
         Turner, N. H., Hartkopf, W. I., \& Bagnuolo, W. G., Jr. 
         2000, \aj, 119, 2403 
\bibitem[Turner \& Forbes(1982)]{turner1982}
         Turner, D. G., \& Forbes, D. 1982, \pasp, 94, 789
\bibitem[Turner et al.(2008)]{tur08}
         Turner, N. H., ten Brummelaar, T. A., \& Mason, B. D. 2008, 
         BAAS, 40, 208
\bibitem[Valdes et al.(2004)]{val04}
         Valdes, F., Gupta, R., Rose, J. A., Singh, H. P., \& Bell, D. J.
         2004, \apjs, 152, 251
\bibitem[van Belle et al.(1999)]{van99}
         van Belle, G. T., et al. 1999, \aj, 117, 521
\bibitem[van Leeuwen(2007)]{van07}
         van Leeuwen, F. 2007, \aap, 474, 653
\bibitem[Walborn(1972)]{wal72}
         Walborn, N. R. 1972, \aj, 77, 312
\bibitem[Walborn \& Fitzpatrick(1990)]{walborn1990}
         Walborn, N.R., \& Fitzpatrick, E.L. 1990, \pasp, 102, 379
\bibitem[Wegner(1994)]{weg94} 
         Wegner, W. 1994, \mnras, 270, 229
\bibitem[Zacharias et al.(2010)]{zac09}
         Zacharias, N., et al. 2010, AJ, 139, 2184

\end{thebibliography}
\end{document}